# On the linear operation of cloned dynamical systems and its Lyapunov exponents

Wang Pengfei[1,2]

1 Center for Monsoon System Research, Institute of Atmospheric Physics, Chinese Academy of Sciences, Beijing 100190

2 State Key Laboratory of Numerical Modeling for Atmospheric Sciences and Geophysical Fluid Dynamics, Institute of Atmospheric Physics, Chinese Academy of Sciences, Beijing 100029

**Abstract** The cloned dynamical system theory is introduced and the Lyapunov exponents of this system are qualitatively proven to be same as the original dynamical system. This property indicates that these two systems have the same error propagation speed in the phase space, and thus we can interpret the phenomenon as why the ensemble mean method sometimes is not effective.

## 1 Introduction

We consider a dynamical system to satisfy

$$\dot{X} = f(X), \text{ where } X = (X_1, X_2, \cdots, X_m)^T, \quad f = (f_1, f_2, \cdots, f_m)^T, \tag{1}$$

and we define $T_X^t = matrix\left(\dfrac{\partial f_i^t}{\partial X_j}\right)$, $T_X^{t*}$ is the adjoint matrix of $T_X^t$.

The multiplicative ergodic theorem tells us that if the following limits exists (Oseledec 1968, Ruelle 1979, Eckmann and Ruelle 1985):

$$\lim_{t \to \infty} \left(T_X^{t*} T_X^t\right)^{\frac{1}{2t}} = \Lambda_X, \tag{2}$$

where $\Lambda_X$ is a matrix, then the Lyapunov exponents can be defined as:

$$\lim_{t \to \infty} \frac{1}{t} \ln \left\| T_X^t u \right\| = \lambda^{(i)}, \quad u \in E_X^{(i)} \backslash E_X^{(i+1)}, \tag{3}$$

where $\lambda^{(1)} > \lambda^{(2)} > \cdots \lambda^{(k)}$, is the logarithm of the eigenvalue of $\Lambda_X$.

Correspondence author: wpf@mail.iap.ac.cn

$E_X^{(i)}$ is the subspace of $R^m$ corresponding to the eigenvalues $\leq \exp \lambda^{(i)}$.

## 2 the independent and associated cloned dynamical systems

**Theorem 1:** We consider an independent cloned dynamical system defined by

$$\begin{cases} \dot{x}_1 = f(x_1) \\ \dot{x}_2 = f(x_2) \\ \vdots \\ \dot{x}_n = f(x_n) \end{cases} \quad (4)$$

Where $x_i = (x_{i,1}, x_{i,2}, \cdots, x_{i,m})^T$ and $f$ are the same as in equation (1). Each sub-systems in equation (4) have different initial values, $x_i(0)$. Since they following the same dynamical function, $f$, we call it an $n$-th cloned dynamical system. The Lyapunov exponents of equation (4) are the same as in equation (1) but the number of Lyapunov exponents is $nk$.

**Proof:**

Because each $\dot{x}_i = f(x_i)$ in equation (4) is independent of other components and can be looked upon as an independent and self-organized dynamical system, each of the sub-systems is $m$-dimensional and in total we have $n \times m$ dimensions.

The Jacobian matrix for equation (4) is

$$J = \begin{bmatrix} J_1 & 0 & 0 & 0 \\ 0 & \ddots & 0 & 0 \\ 0 & 0 & J_{n-1} & 0 \\ 0 & \cdots & 0 & J_n \end{bmatrix}, \quad (5)$$

where $J_i = T_X^t = matrix\left(\dfrac{\partial f_i^t}{\partial X_j}\right)$. When we obtain the limit of this matrix, the value will be determined by the main diagonal matrix:

$$\lim_{t \to \infty}(J^*J)^{\frac{1}{2t}} = \begin{bmatrix} \Lambda_1 & 0 & 0 & 0 \\ 0 & \ddots & 0 & 0 \\ 0 & 0 & \Lambda_{n-1} & 0 \\ 0 & \cdots & 0 & \Lambda_n \end{bmatrix}. \qquad (6)$$

The ergodic property guarantees that $\Lambda_i = \Lambda_X$, moreover, the Lyapunov exponents of each system can be defined by equation (3), and the $n$-th subsystem together has $nk$ exponents.

**Theorem2:** We consider a linear system associated with a cloned dynamical system defined by $y = \sum_{i=1}^{n} c_i x_i$, and $c_i \neq 0$, where $x_i$ is same as equation (4), ie.

$$\begin{cases} \dot{x}_1 = f(x_1) \\ \dot{x}_2 = f(x_2) \\ \vdots \\ \dot{x}_{n-1} = f(x_{n-1}) \\ \dot{x}_n = f(x_n) \\ y = \sum_{i=1}^{n} c_i x_i = \sum_{i=1}^{n} c_i f(x_i) \equiv F(x_1, x_2, \cdots, x_n) \end{cases} \qquad (7)$$

Transforming the variable in the sub-equation $\dot{x}_n = f(x_n)$ by $x_n = \frac{1}{c_n} y - \sum_{i=1}^{n-1} \frac{c_i}{c_n} x_i$ will obtains the equation $\dot{y} = g(x_1, x_2, \cdots, x_{n-1}, y)$.

$$\begin{cases} \dot{x}_1 = f(x_1) \\ \dot{x}_2 = f(x_2) \\ \vdots \\ \vdots \\ \dot{x}_{n-1} = f(x_{n-1}) \\ \dot{y} = g(x_1, x_2, \cdots, x_{n-1}, y) = \sum_{i=1}^{n-1} c_i f(x_i) + c_n f\left(\frac{1}{c_n} y - \sum_{i=1}^{n-1} \frac{c_i}{c_n} x_i\right) \end{cases}, \qquad (8)$$

and the Lyapunov exponents of $\dot{y} = g(x_1, x_2, \cdots, x_n, y)$ are the same as in equation (1).

**Proof:**

the Jacobian matrix for equation (8) is $J = \begin{bmatrix} J_1 & 0 & 0 & 0 \\ 0 & \ddots & 0 & 0 \\ 0 & 0 & J_{n-1} & 0 \\ R_1 & \cdots & R_{n-1} & J_n \end{bmatrix}$, where

$J_i = T_X^t = matrix\left(\dfrac{\partial f_i^t}{\partial X_j}\right)$ and $R_i = matrix\left(\dfrac{\partial g_i^t}{\partial x_j}\right)$. It has different values to equation (5) in the last $m$ rows, but when we obtain the limits,

$$\lim_{t \to \infty}(J^*J)^{\frac{1}{2t}} = \begin{bmatrix} \Lambda_1 & 0 & 0 & 0 \\ 0 & \ddots & 0 & 0 \\ 0 & 0 & \Lambda_{n-1} & 0 \\ 0 & \cdots & 0 & \Lambda_n \end{bmatrix} \tag{9}$$

it will be determined by the main diagonal matrix. The ergodic property guarantees that $\Lambda_i = \Lambda_X$, so each Lyapunov exponents is the same as in equation (1), including the maximal Lyapunov exponents.

## 3. Conclusion and discussion

Following theorem 2, and taking $c_i = \dfrac{1}{n}$, thus $y = \dfrac{1}{n}\sum_{i=1}^{n}x_i$ is the ensemble mean system of $x_i$, and its Lyapunov exponents are the same as in equation (1). Meanwhile, the maximal Lyapunov exponents for the equation $\dot{y} = g(x_1, x_2, \cdots, x_n, y)$ are $\lambda^{(1)}$.

Because the ensemble mean dynamical system has the same maximal Lyapunov exponents, thus the prediction time is thus $T_p \sim \dfrac{1}{\lambda_{max}}\ln(\dfrac{\Delta}{\delta_0})$ (where $\Delta$ is a given tolerance) and will be same as the original dynamical system in equation (1).

This discussion is under the assumption of linear error propagation, when considering nonlinear error propagation different dynamics will be obtained. However, when $\delta_0 \to 0$, the difference of $T_p$ between linear and nonlinear cases is not significant(Ding and Li 2007).

**Acknowledgements:** This research was jointly supported by the National Natural Sciences Foundation of China (41375112) and the National Basic Research Program of China (2011CB309704).